\documentclass[aps,prl,twocolumn,superscriptaddress, nofootinbib]{revtex4-2}
\usepackage{xcolor}
\usepackage{multirow}
\usepackage{array}
\usepackage{makecell} 
\usepackage{braket}
\usepackage{graphicx}
\usepackage{amsmath}
\usepackage{amssymb}
\usepackage{hyperref}
\usepackage{enumitem}
\usepackage{float}
\usepackage{booktabs}
\usepackage{tabularx}
\usepackage{array}
\usepackage{makecell}
\setcellgapes{1.5pt}
\newcolumntype{Y}{>{\centering\arraybackslash}X}
\makegapedcells     

\begin{document}
\title{Classical and quantum mechanics across representations:\\ an operational reading of the Wigner–Weyl correspondence}
\author{Samuel Schlegel}
\affiliation{University of Vienna, Faculty of Physics, Vienna Center for
Quantum Science and Technology, Boltzmanngasse 5,
Vienna 1090, Austria}
\author{Borivoje Daki\'c}
\affiliation{University of Vienna, Faculty of Physics, Vienna Center for
Quantum Science and Technology, Boltzmanngasse 5,
Vienna 1090, Austria}
\affiliation{Institute for Quantum Optics and Quantum Information (IQOQI),
Austrian Academy of Sciences, Boltzmanngasse 3, Vienna 1090,
Austria}
\author{Flavio Del Santo}
\affiliation{Constructor University, Bremen, Germany}
\affiliation{University of Geneva, Group of Applied Physics, Rue de l'\'Ecole-de-M\'edecine 21, Geneva 1211, Switzerland}

\begin{abstract}
The physical content of a theory is not intrinsically tied to any single mathematical formalism. Both classical and quantum mechanics admit equivalent representations, notably in phase space and in Hilbert space, related by the Wigner--Weyl correspondence. While this correspondence has long been studied in mathematical physics, its foundational and operational implications are often left implicit. Here we give a systematic account of what changes---and what does not---when classical and quantum theories are expressed in each other's native language. This representational viewpoint separates artifacts (such as the appearance of non-positivity or negativity under certain maps) from robust structural distinctions that persist across representations, in particular noncommutativity and its $\hbar$-dependent $\star$--deformation of the classical algebra. We develop the comparison at the level of states, kinematics, and dynamics, and extend it to measurement by formulating both outcome statistics and state-update rules within the same framework.
\end{abstract}

\maketitle

\section{Introduction}
A recurrent lesson in theoretical physics is that the same physical theory can be expressed in different mathematical languages. Yet, in textbook presentations, classical and quantum mechanics are often implicitly identified with their most familiar representations: classical mechanics with phase space (PS), and quantum mechanics with Hilbert space (HS). This association encourages the impression that the classical--quantum divide is, at least in part, a divide between formalisms. A formalism does not change what a theory predicts, but it does shape how the theory is understood, including which of its features appear fundamental. Already in the early development of quantum theory, Moyal articulated this discomfort: “\textit{One is led to wonder whether this formalism does not disguise what is an essentially statistical theory, and whether a reformulation of the principles of quantum mechanics in purely statistical terms would not be worth while in affording us a deeper insight
into the meaning of the theory}”~\cite{moyal_quantum_1949}. In fact, Wigner, Weyl, and others~\cite{wigner_quantum_1932, weyl_quantenmechanik_1927, groenewold_principles_1946, van_hove_sur_1951, baker_formulation_1958} established early on that quantum mechanics admits a phase-space representation, and that phase-space structures can, conversely, be encoded in Hilbert-space terms. This converse direction has historically drawn far less attention than the quantum to phase-space map, the principal exception being the algebraic phase-space program of Bohm and Hiley~\cite{bohm_quantum_1981}, even though it is essential for a symmetric comparison. The broader consequences of treating these representations on an equal footing, however, have often been explored mainly from the perspective of mathematical physics, with less emphasis on the operational and foundational reading of what the translations do (and do not) change~\cite{curtright_quantum_2012}.

In parallel, a sustained effort in recent years has aimed at sharpening the classical--quantum boundary by pushing ``classical'' models to emulate hallmark quantum features. Examples include epistemic restrictions designed to mimic uncertainty~\cite{bartlett_reconstruction_2012, spekkens_defense_2007}, proposals of fundamental indeterminacy in otherwise classical settings~\cite{santo_physics_2019, santo_which_2025}, or limitations of operational access imposed on observers~\cite{schlegel_entanglement_2025}. These programs underscore an important point: many striking ``quantum-like'' features depend on what is operationally accessible and on how a theory is represented, rather than on a simple ontological split between ``classical'' and ``quantum.''

The goal of the present work is to push this representational comparison further by systematically translating each theory into the language of the other and tracking which distinctions survive the change of representation. We use the Wigner--Weyl (WW) correspondence~\cite{gerry_introductory_2023, case_wigner_2008, hillery_distribution_1984} to map classical PS distributions to HS operators (not necessarily positive) and, conversely, quantum density operators to their PS representatives (Wigner functions). While other Hilbert-space embeddings of classical mechanics exist (e.g., Koopman--von Neumann~\cite{mauro_topics_2003}), these proceed from a \emph{commutative} algebra of observables embedded in Hilbert space, building in commutativity from the beginning. Related approaches, however, show that operator and Hilbert-space methods, and even certain noncommutative structures \cite{morgan_algebraic_2020}, along with hallmark features such as complementarity \cite{galazo_classical_2020}, can emerge in otherwise classical settings, so that none of these is by itself uniquely quantum. The WW framework is instead particularly well suited for a two-way comparison precisely because  the noncommutative composition law is retained at the level of symbols, so the principal distinction is relocated to the admissible states and the dynamics rather than assumed away in the kinematics.

In essence the formalism provides a two way translation between phase space and Hilbert space descriptions: quantum states can be represented in phase-space form and classical phase-space descriptions can be represented in Hilbert-space form. These transitions are exact on the level of representation, but they do not, in general, preserve all the admissibility constraints in the same way. As a result some familiar indicators, such as negativity appearing in one representation, can reflect the representational choice rather than a representation-independent physical boundary.

The aim of this work is to use this two-way translation as an operational comparison tool and to separate representational artifacts from structural differences between classical and quantum theories. Our main claim is that the robust boundary is not tied to a preferred language (phase-space vs Hilbert-space), but to noncommutative structure and its operational consequences. Put differently, we ask: after translating both theories into each other’s native language, which differences remain genuinely physical?

The goal of this paper is twofold. On the one hand, we review the main features of the Wigner–Weyl (WW) formalism, originally developed in the context of mathematical physics, and reinterpret them from an operational perspective. This viewpoint brings to light a number of conceptual and foundational aspects of the WW correspondence that, in our opinion, have received comparatively little attention in the literature. On the other hand, we extend the WW framework with several original contributions. We provide a systematic analysis of which differences between the classical and quantum descriptions are merely artifacts of representation and which instead reflect genuine structural distinctions, carrying out this comparison uniformly at the level of states, dynamics, and measurements. We formulate outcome probabilities and conditional state-update rules within a common operational framework, thereby allowing a direct comparison between classical and quantum theories. This unified perspective clarifies which apparent signatures of nonclassicality are representation-dependent and which instead originate from the underlying noncommutative algebraic structure, making them representation-independent. Finally, we show how recent results on entanglement naturally fit within this same operational hierarchy.

While some of the ideas of this work were introduced in Ref. \cite{schlegel_entanglement_2025}, that paper is a focused study of bipartite continuous-variable correlations and establishes the entanglement hierarchy that we review in Sec.~\ref{sec:entanglement}. The present work is instead concerned with a systematic comparison of the two formalisms. Its contribution is to carry the same operational reading uniformly through states, kinematics, dynamics, measurement and entanglement, such that the question of what is representational and what is structural can be asked and answered in the same terms. This is also what allows the comparison to be extended to measurement, which is usually left out of such treatments, by expressing outcome statistics and conditional state updates for both theories in a common language and locating the difference in the update rule rather than in the outcome probabilities.

After formally introducing the WW maps, we will discuss the admissibility constraints on phase-space functions, including known necessary-and-sufficient criteria (the Kastler–Loupias–Miracle-Sole (KLM) / quantum Bochner-type conditions~\cite{Narcowich1989, kastler1965c, Loupias_Miracle_1966, Bochner1933}) and practically useful special cases (e.g. Gaussian states) \cite{narcowich_necessary_1986, Narcowich1989}. We will find that the classical statespace $\mathcal{C}$ and the quantum statespace $\mathcal{Q}$ form distinct partially overlapping sets, with an intersection that admits both a classical and a quantum description. To further find real physical differences between the two theories however, we will also have to look beyond the statespace and consider the underlying differences in the kinematics and the dynamics, in both PS and HS. In doing so, we immediately obtain the generators for time evolution in the form of the Poisson and the Moyal bracket, which can then be translated into an HS language, revealing subtle structural differences between the two evolutions. Furthermore, the quantum measurement process, including both the calculation of probabilities, as well as the state update, will be compared against its classical counterpart. Once translated into the same language, we will find great similarities in the calculation of outcome probabilities, with the only exception being emerging negativity inherently tied to this correspondence. If one were to restrict the states and the measurements an observer can perform (e.g.\ to jointly measurable phase-space observables), these issues can be avoided. Nevertheless, there will be a difference in the state update rule in PS that can be used to distinguish both processes. Finally, we will also briefly review the emerging entanglement structure of Ref. \cite{schlegel_entanglement_2025}, to complete the comparison in Table \ref{tab:compare}.

\section{Representational Translations}
The Wigner--Weyl correspondence provides an explicit change of representational language between HS and PS. At the level of symbols/operators, the translation is invertible, so no information is lost. However, as we make precise below once the transforms are defined, the map does not preserve the positivity constraints that define the \emph{physical} state spaces and this mismatch is the first place where representational artifacts such as negativity can appear. Yet we will find that both HS and PS representations are not intrinsically tied to either theory. This section fixes notation and makes precise the distinction between formal translations and admissible state sets, which will be crucial for the kinematic, dynamical, and measurement comparisons below.

Quantum states are customarily represented in complex Hilbert space $\mathcal{H}$, where pure states are represented by complex vectors $\ket{\psi}$ and mixed states are represented by positive trace 1 operators $\hat{\rho} = \sum_i \omega_i \ket{\psi_i} \bra{\psi_i}$. With the knowledge of the state, expectation values of an operator $\hat{O}$ can be computed via $\langle \hat{O} \rangle = \text{Tr}[\hat{\rho} \hat{O}]$. To then transform a quantum state to its corresponding quasi--probability distribution in PS, we make use of the Wigner transform 
\begin{equation}
\label{eq:Wigner_function}
    W(q,p) = \frac{1}{2\pi\hbar}\int_{-\infty}^\infty dy \, e^{-\frac{i}{\hbar}py}\bra{q+\frac{y}{2}} \hat{\rho} \ket{q-\frac{y}{2}},
\end{equation}
where the resulting Wigner--function $W$ is real--valued but can take negative values \cite{case_wigner_2008}. The transformation admits an inverse at the level of operators/symbols, which can be used to represent classical phase-space functions as Hilbert-space operators.

These classical states with $n$ degrees of freedom, are typically represented by a probability distribution $f(q,p)$ on phase space $\Gamma \cong \mathbb{R}^{2n}$, where deterministic (pure) states correspond to Dirac delta functions $f(q,p) = \delta(q-q_0)\delta(p-p_0)$, while statistical ensembles are described by smooth probability densities (mixed states). Observables can then be represented by real-valued functions $O(q,p)$ on $\Gamma$, for which the expectation values can be computed via $\langle O \rangle = \int dq dp f(q,p) O(q,p)$. To obtain the Hilbert space counterpart of these well-behaved PS distributions, we require the Weyl quantization map $\Phi$ \cite{weyl_quantenmechanik_1927}, which is defined by
\begin{equation}
\label{eq:Weyl_map}
    \hat{\rho}_f = \Phi[f] = \int_{\mathbb{R}^{2n}} dq \, dp \, f(q,p) \hat{\Delta}(q,p),
\end{equation}
where the Stratonovich--Weyl kernel \cite{stratonovich1957distributions} is given by
\begin{equation}
    \hat{\Delta}(q,p) = \int_{\mathbb{R}} ds\, e^{\frac{i}{\hbar}ps} \ket{q+\frac{s}{2}}\bra{q-\frac{s}{2}}.
\end{equation}
Using this Weyl quantization allows us to associate any well-behaved PS distribution $f$ with its HS operator counterpart $\hat{\rho}_f$. Like for the Wigner transform, this mapping does not preserve positivity, i.e., a positive classical PS distribution can potentially be mapped to a non--positive semidefinite (pseudo--)density operator. Consequently, the translation is formally exact while the admissible sets differ, and positivity fails in \emph{both} directions: a positive classical density $f\ge 0$ can map to an operator $\hat\rho_f\not\ge 0$, while a positive operator $\hat\rho\ge 0$ can map to a Wigner function that is pointwise negative. Quantum states are thus positive operators whose Wigner functions need not be pointwise positive, whereas classical phase-space densities are positive functions whose Hilbert-space images $\hat{\rho}_f$ need not be positive operators. Much of what is often advertised as ``nonclassicality'' (e.g.\ negativity) therefore depends on which representation one insists on.

In what follows we assign states the normalization of Eq. \eqref{eq:Wigner_function} and observables the bare Weyl symbol. With this convention the trace relation below holds as written whenever one argument is a (pseudo-)density and the other an observable, which is the only case we use. 
Thus one can translate the calculation of expectation values from one formalism to the other, via the trace relation \cite{hillery_distribution_1984}
\begin{equation}
\label{eq:trace}
    \text{Tr}(\hat{A} \hat{B}) = \int dq \, dp \, A_W(q,p) B_W(q,p),
\end{equation}
where subscript $W$ denotes the corresponding symbol. 

\begin{figure}[h]
    \centering
    \includegraphics[width=0.9\linewidth]{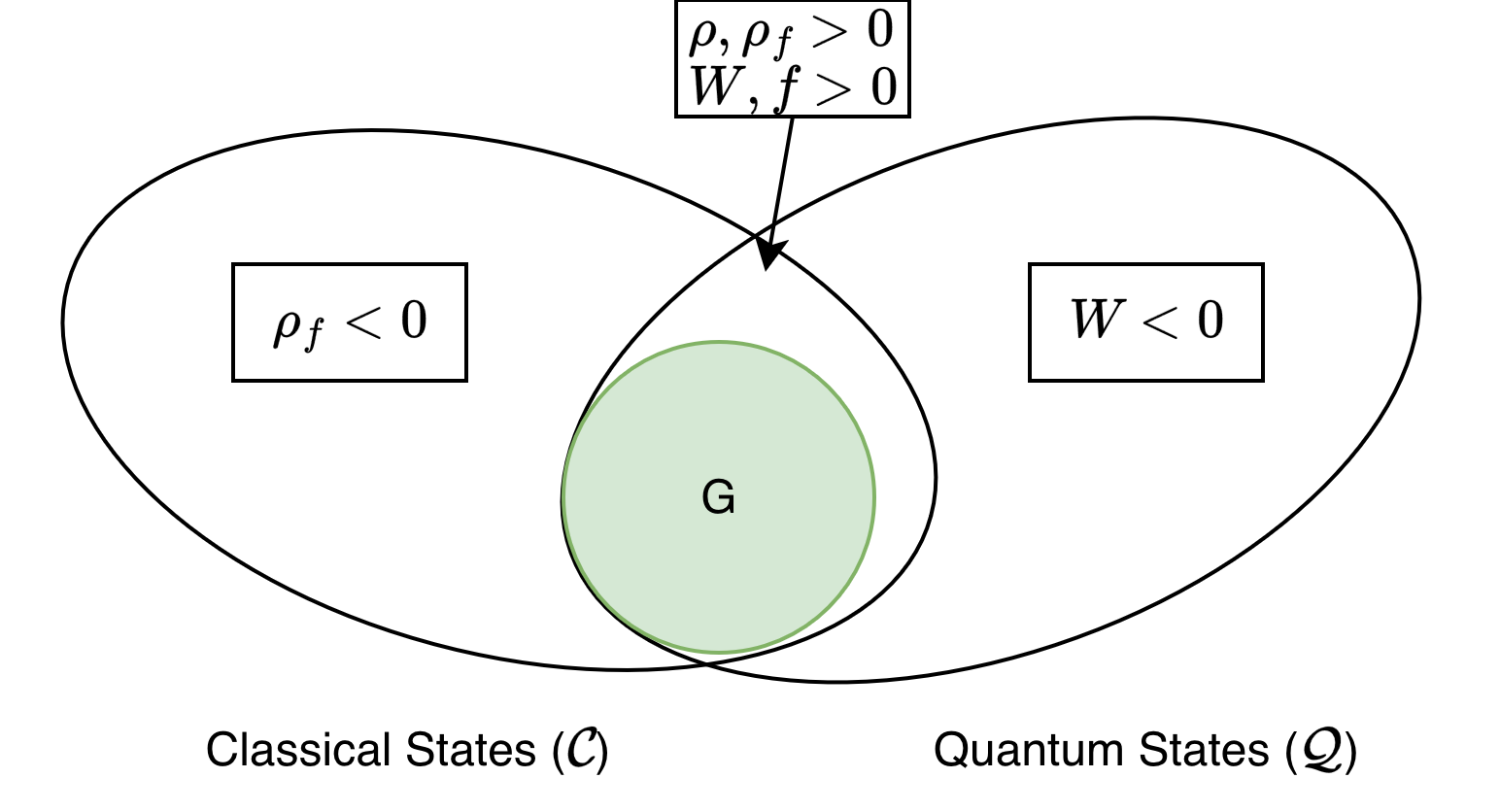}
    \caption{Classical and quantum state spaces are partially overlapping sets when considered under the WW correspondence. The left region contains classical densities $f\ge0$ whose Hilbert-space images $\hat\rho_f$ are not positive operators, the right region contains positive density operators $\hat\rho$ whose Wigner functions take negative values somewhere, and the intersection contains the states that meet both positivity requirements at once. Gaussian quantum states $G$ can be found within the intersection, by Hudson's theorem~\cite{hudson_when_1974}.}
    \label{fig:WW}
\end{figure}

Just this mapping of states already reveals a rich set-theoretic structure. The physically admissible classical and quantum state spaces become distinct, partially overlapping sets when viewed through the WW translations \cite{schlegel_entanglement_2025}. Importantly, (non)positivity is not preserved by the correspondence, so negativity should be read as a \emph{representation-relative witness} rather than an intrinsic label. For example, a Wigner function that takes negative values rules out an interpretation as an ordinary classical phase-space probability density within this representation and is therefore often used as a witness of nonclassicality \cite{kenfack_negativity_2004, mandilara_extending_2009}. Conversely, the Hilbert-space image $\hat{\rho}_f$ of a perfectly classical density $f$ need not be positive, which means it cannot be regarded as a quantum state even though it is a valid WW representative of a classical distribution. More formally, we can define the admissible set of quantum states $\mathcal{Q}$ as 
\begin{equation}
\label{eq:def_Q}
    \mathcal{Q} = \{ \hat{\rho} \;| \;\hat{\rho} \geq 0, \text{Tr}[\hat{\rho}] = 1\},
\end{equation}
and the set of classical states via
\begin{equation}
\label{eq:def_C}
    \mathcal{C} = \{ \hat \rho_f \;| \;\text{Tr}[\hat{\rho}_f] = 1, f(q,p) \geq 0 \;\forall \;(q,p)\}
\end{equation}
This structure is summarized schematically in Fig.~\ref{fig:WW} and in the first row of Table~\ref{tab:compare}. To make the nature of the classical and quantum state sets more concrete, consider the following examples. The classical pure state $f=\delta(q-q_0)\delta(p-p_0)$ has Weyl image $\hat\rho_f=\hat\Delta(q_0,p_0)$, i.e.\ the Stratonovich--Weyl kernel itself, which is proportional to the parity operator about the point $(q_0,p_0)$ and therefore carries eigenvalues of both signs. A perfectly ordinary classical state is therefore mapped outside $\mathcal{Q}$. On the quantum side, the Fock state $\ket{1}$ is a legitimate density operator whose Wigner function takes the value $W(0,0)=-1/(\pi\hbar)$ at the origin, so it admits no reading as a classical phase-space density. At the same time, negativity is not inevitable, as we can see for Gaussian quantum states ($G$ in Fig. \ref{fig:WW}), for instance, which have everywhere nonnegative Wigner functions \cite{hudson_when_1974}, yielding a nontrivial overlap region in which the two descriptions agree on positivity. Thus, the WW maps provide a formally invertible translation between descriptions, while the \emph{admissible} state sets are singled out by different positivity constraints; in this sense, neither phase-space nor Hilbert-space descriptions are intrinsically ``classical'' or ``quantum.''

Since the WW maps are invertible but do not always preserve positivity, one may ask for criteria that decide whether a given phase space distribution $W(q,p)$ corresponds to the Wigner function of some positive density operator $\hat \rho \geq 0$. A complete answer is given by the KLM conditions \cite{Narcowich1989, CorderoDeGossonNicola2019, DiasPrata2005}, using the characteristic function $\widetilde{W}(\xi)$. Here $\widetilde{W}(\xi)$ is the (symplectic) Fourier transform of $W$, the PS analogue of the characteristic function of a probability distribution. Classically, Bochner's theorem states that a continuous function normalized to 1 at the origin is the characteristic function of a genuine probability density if and only if it is of \emph{positive type}, i.e. the kernel $\widetilde{W}(\xi_j-\xi_k)$ is positive semidefinite on every finite set $\{\xi_j\}$. The KLM conditions are the quantum deformation of this criterion. Positive--definiteness is replaced by $\hbar$--\emph{positive type}, in which each entry carries an additional phase $\exp \left(  \frac{i}{2\hbar}\sigma(\xi_j, \xi_k)\right)$.  More concretely, $W$ is the Wigner function of a legitimate state $\hat \rho \geq 0$ iff $\widetilde{W}(0) = 1$, $\widetilde{W}$ is continuous, and the matrix
\begin{equation}
    M_{jk} = \widetilde{W}(\xi_j - \xi_k) \exp \left( \frac{i}{2\hbar} \sigma(\xi_j, \xi_k) \right),
\end{equation}
is positive semidefinite where $\sigma$ denotes the symplectic form. This is a quantum analogue of Bochner's theorem, which is recovered when $\hbar \rightarrow 0$. While these conditions are not always convenient computationally, they make the admissible subset in Fig. \ref{fig:WW} mathematically sharp, and they admit certain special cases. Most notably for Gaussian states, admissibility reduces to the Robertson--Schrödinger uncertainty constraint on the covariance matrix~\cite{serafini_quantum_2023, weedbrook_gaussian_2012}.

It is useful to fix the terminology at this point. By a \emph{classical phase-space description} we mean a nonnegative probability density on $\Gamma$ together with the machinery that comes with it, i.e. Liouville evolution, response functions as measurements and Bayes' rule as the state update. By \emph{Wigner positivity} we mean pointwise nonnegativity of $W(q,p)$. By \emph{classicality} we always mean the operational property of being reproducible by a classical phase-space description under a stated set of accessible measurements, and not an ontological attribution. Wigner positivity is necessary for a classical phase-space account of the quadrature statistics of a quantum state, but it is not sufficient for classicality in the above sense, because the two theories can still be told apart by the dynamics under non-quadratic Hamiltonians, as in Sec.~\ref{sec:time_evolution}, and by the state update, as in Sec.~\ref{sec:measurement}. Conversely, a state can be classical in origin and still fail to be a quantum state at all.

So far the WW correspondence has given us an invertible translation that allows us to map from one representation to the other, while positivity constraints pick out different physical subsets. To now locate a representation--independent distinction, we shall look at what the WW--correspondence does \emph{not} preserve structurally. We will find that multiplication in one language is different in the other, and that mismatch is where noncommutativity enters.

\newcommand{\cbox}[1]{\parbox[c]{.21\textwidth}{\centering #1}}
\begin{table*}[t]
\caption{\label{tab:compare}Comparison of states, evolution, kinematics, measurement, and entanglement
in classical and quantum mechanics, across Hilbert and phase--space formalisms.}
\renewcommand{\arraystretch}{2}
\setlength{\tabcolsep}{4pt}
\scriptsize
\centering
\begin{tabular}{lcccc}
\toprule
 & \multicolumn{2}{c}{\textbf{Phase space}} & \multicolumn{2}{c}{\textbf{Hilbert space}} \\
\cmidrule(lr){2-3}\cmidrule(lr){4-5}
 & \textbf{Classical} & \textbf{Quantum} & \textbf{Classical} & \textbf{Quantum} \\
\midrule
State
 & \cbox{$f(q,p)\ge 0$}
 & \cbox{$W(q,p)$ may be\\ negative}
 & \cbox{$\hat\rho_f$ need not be\\positive}
 & $\hat\rho\ge 0$ \\
 Kinematics
 & $\{q,p\}_{\mathrm{PB}}=1$
 & $\{\cdot,\cdot\}_{\mathrm{MB}}=\{\cdot,\cdot\}_{\mathrm{PB}}+O(\hbar^2)$
 & $[\hat q,\hat p]=i\hbar$
 & $[\hat q,\hat p]=i\hbar$ \\
Evolution
 & $\partial_t f=\{H,f\}_{\mathrm{PB}}$
 & $\partial_t W=\{H,W\}_{\mathrm{MB}}$
 & BH
 & von Neumann \\
Measurement
 & \cbox{Bayes conditioning\\(likelihood reweighting)}
 & \cbox{Overlap rule\\ $\star$--deformed update}
 & \cbox{formal trace rule\\ (positivity not preserved)}
 & \cbox{Born rule:\\projective/POVM} \\
Entanglement~\cite{schlegel_entanglement_2025}
 & \cbox{probabilistic\\correlations only}
 & \cbox{genuine\\entanglement}
 & \cbox{representational\\``entanglement''}
 & \cbox{operational\\quantum\\entanglement} \\
\bottomrule
\end{tabular}
\end{table*}

\section{Kinematics}
Before comparing algebras, it is worth separating two notions that the translation tends to conflate: \emph{variables} and \emph{observables}. Classically, the canonical pair $(q,p)$ are ordinary commuting phase-space coordinates; under Weyl quantization their images $(\hat q,\hat p)$ obey $[\hat q,\hat p]=i\hbar$. One might read this acquired noncommutativity as evidence that the embedded classical system is somehow ``intrinsically quantum''. It is not. The commutator here is a property of the representation \emph{map} applied to the kinematic variables, not of the classical state of affairs being described. Concretely, the embedding changes no classical prediction, because the operator
image $\hat\rho_f=\Phi[f]$ of a classical density has Wigner function equal to $f$ itself, so every quadrature statistic is exactly the classical one, even though the coordinate images $\hat q,\hat p$ no longer commute. The commutator acquires operational meaning only when something is sensitive to it, i.e. when positivity forbids arbitrarily sharp joint localization (the uncertainty bound below, Eq.~\eqref{eq:RS_matrix}) or when the $\star$-product enters a measurement update (a sharp position readout disturbing momentum, Sec.~\ref{sec:measurement}). Until then, $[\hat q,\hat p]=i\hbar$ merely records how one chose to write classical functions as operators. Thus what distinguishes the theories is not whether one can write noncommuting symbols, but whether the admissible \emph{states} (through positivity) and the operationally accessible measurements (through the $\star$--deformed composition) are sensitive to that structure (see Sec.~\ref{sec:measurement}). We introduce the Poisson and Moyal brackets here as the algebraic objects induced by the two composition laws, while their role as generators of motion is taken up in the next section.

With transformations and state spaces in hand, we can now ask what algebraic structure needs to be changed under this mapping to make a theory consistent in both representations. The key point is not whether we can translate symbols back and forth, but whether the translation respects the \textit{product structure} that underlies kinematics and dynamics. In the following, we will motivate the emergence of the Moyal $\star$--product via the Groenewold--van Hove obstruction and find how the mapping can be made to satisfy certain quantization properties. This will lead us to find the structural differences in the kinematic relations in PS. 

Notice that the Weyl--transform in Eq. \eqref{eq:Weyl_map} ensures that the constant function $1$ is mapped to the identity operator $\mathbb{I}$, and polynomials are mapped to their symmetrized counterparts, e.g., $\Phi[qp] = (\hat{q}\hat{p} + \hat{p}\hat{q})/2$ \cite{fujii_new_2004, cahill_ordered_1969}. Despite that, this quantization procedure cannot yet serve as a fully consistent quantization map \cite{carosso_quantization_2022}, due to the Groenewold-van Hove theorem \cite{groenewold_principles_1946, van_hove_sur_1951}, which shows that a quantization map $Q$ needs to satisfy: (i) normalization $Q(1) = \mathbb{I}$, (ii) canonical variables $Q(q_i) = \hat{q}_i$ and $Q(p_i) = \hat{p}_i$, and (iii) the Dirac condition $Q(\{f,g\}) = \frac{1}{i\hbar}[Q(f), Q(g)]$, which maps the Poisson brackets to a commutator structure. This theorem states that there exists no quantization map that satisfies these properties, i.e., preserves the Poisson algebra exactly. Particularly, problems arise when considering polynomials of third order 
\begin{equation}
    i \hbar Q(\{q^2p, qp^2\}) \neq [Q(q^2p), Q(qp^2)],
\end{equation}
which will have a reoccurrence when we consider time evolution in a later section, where third order potentials lead to a divergence between classical and quantum evolution. Note that up to quadratic order, there is no need to deform the PS algebra, as all of the above properties are satisfied. Since an exact correspondence is not possible to all orders, we have to relax condition (iii) and introduce a deformation of the Poisson structure in the form of the Moyal bracket \cite{moyal_quantum_1949, bayen_deformation_1978_I, bayen_deformation_1978_II}
\begin{equation}
\label{eq:MB}
\{f,g\}_\text{MB} = \frac{1}{i\hbar}( f \star g - g \star f) = \{f,g\} + \mathcal{O}(\hbar^2),
\end{equation} 
where the Moyal $\star$--product is defined as 
\begin{equation}
    f \star g = f(q,p) \exp \left[ \frac{i \hbar}{2}\left(  \overleftarrow{\partial_q} \overrightarrow{\partial_p} - \overleftarrow{\partial_p} \overrightarrow{\partial_q})\right) \right] g(q,p),
\end{equation}
and $\overleftarrow{\partial}/\overrightarrow{\partial}$ acts to the left/right, respectively. Operationally, the $\star$-product is just the phase-space encoding of operator multiplication and tells us how to multiply Weyl symbols so that the result matches multiplying the corresponding Hilbert-space operators. This new product is also used to encapsulate products in Eq. \eqref{eq:trace}, if, for instance, two-point correlators like $\langle \hat A \hat B \rangle \sim \int (A_W \star B_W)W$ are considered, and it generally helps with computing arbitrary products of Wigner transformed operators. It is now also instructive to look at how the Moyal product compares to the regular pointwise product in phase space. Expanding the exponential makes the deformation explicit
\begin{equation}
\label{eq:star_expansion}
    f \star g = f g + \frac{i\hbar}{2} \{f,g\} + \mathcal{O}(\hbar^2),
\end{equation}
so the antisymmetric part reproduces the Poisson bracket at leading order, while higher even powers in $\hbar$ generate genuinely quantum corrections. The Moyal bracket is then precisely the bracket induced by this deformed product, i.e., the phase-space representative of the commutator. An important consequence of this construction is that classical mechanics can be recovered by taking the limit $\hbar \rightarrow 0$, for which the Moyal bracket reduces to its Poisson counterpart and the Moyal product reduces to the regular phase space product. Most importantly, the Moyal bracket is precisely the phase-space symbol of a Hilbert-space commutator, a point we make explicit for the canonical pair once the product rule \eqref{eq:Weyl_product_rule} is in hand.

This observation can be sharpened into a clean kinematic comparison. In the WW correspondence, one should not compare \emph{pointwise} multiplication of phase-space functions with operator multiplication; rather, operator multiplication corresponds to the $\star$--product of symbols
\begin{equation}
\label{eq:Weyl_product_rule}
(\hat A \hat B)_W \;=\; A_W \star B_W.
\end{equation}
Applied to the canonical pair, this rule reproduces the commutator directly: $q\star p - p\star q = i\hbar$, i.e.\ $\{q,p\}_{\mathrm{MB}}=1$, whose Hilbert-space image is $[\hat q,\hat p]=i\hbar\,\mathbb{I}$. The canonical commutator is therefore a \emph{consequence} of the deformed product, not an independent input. As a further consequence, commutators in Hilbert space correspond exactly to Moyal brackets in phase space,
\begin{equation}
\label{eq:comm_MB}
\frac{1}{i\hbar}[\hat A,\hat B]_W \;\overset{\mathrm{WW}}{\longleftrightarrow}
\; \{A_W,B_W\}_{\mathrm{MB}},
\end{equation}
while the classical Poisson bracket governs the undeformed (commutative) phase-space algebra. In this sense, the representation-independent kinematic distinction is the underlying algebraic structure. Classical mechanics is governed by a commutative product and the Poisson bracket, whereas quantum mechanics is governed by a noncommutative product whose phase-space representative is the $\star$--product and whose bracket is the Moyal bracket. Furthermore, the canonical relations agree at leading order in $\hbar$, as seen in Eq. \eqref{eq:MB} so the classical kinematics is recovered as the $\hbar\to 0$ limit of the quantum phase-space algebra. Conversely, the Groenewold--van Hove obstruction is precisely the statement that one cannot keep the classical Poisson algebra \emph{and} the classical pointwise product while mapping exactly into operators; the deformation encoded in $\star$ is what makes the correspondence consistent. The representation-independent distinction is the noncommutative composition law. In phase space this appears as the $\star$-product and Moyal bracket, whereas classical mechanics keeps pointwise multiplication and the Poisson bracket.

This vantage point also clarifies the status of the uncertainty relations. Kinematics fixes only the commutator $[\hat q,\hat p]=i\hbar$ (equivalently $\{q,p\}_{\mathrm{MB}}=1$) and by itself this is an algebraic identity and carries no bound on variances. To see what supplies the bound, let $\delta\hat r_i = \hat r_i - \langle \hat r_i\rangle$ with $\hat{\mathbf r}=(\hat q,\hat p)$, and consider $\hat X = \mathbf{u}\cdot\delta\hat{\mathbf r}$ for an arbitrary complex vector $\mathbf{u}$. Positivity of the state guarantees
\begin{equation}
\langle \hat X^\dagger \hat X\rangle = \mathrm{Tr}\!\left[\hat\rho\,\hat X^\dagger \hat X\right] \ge 0
\quad \forall\, \mathbf{u}.
\end{equation}
Splitting the second moments into symmetric and antisymmetric parts, $\langle \delta\hat r_i\,\delta\hat r_j\rangle = V_{ij} + \tfrac{i\hbar}{2}\Omega_{ij}$, with covariance matrix $V_{ij}=\tfrac12\langle\{\delta\hat r_i,\delta\hat r_j\}\rangle$ and symplectic form $\Omega = \left(\begin{smallmatrix} 0 & 1\\ -1 & 0\end{smallmatrix}\right)$, the condition for all $\mathbf{u}$ becomes the matrix inequality
\begin{equation}
\label{eq:RS_matrix}
V + \frac{i\hbar}{2}\,\Omega \;\ge\; 0,
\end{equation}
which is precisely the Robertson--Schr\"odinger relation~\cite{serafini_quantum_2023, weedbrook_gaussian_2012}, denoting the same bona-fide constraint that, for Gaussian states, makes the KLM conditions tractable. Taking the determinant gives $\det V \ge \hbar^2/4$, i.e.\ $\Delta q^2\,\Delta p^2 - \mathrm{cov}(q,p)^2 \ge \hbar^2/4$, and hence $\Delta q\,\Delta p \ge \hbar/2$. The commutator enters only through $\Omega$, while the inequality itself is enforced entirely by positivity of $\hat\rho$. Incompatibility is kinematic, but the uncertainty \emph{bound} is a property of the admissible state space. This cleanly separates what the noncommutative product supplies (the bracket) from what positivity supplies (the bound), and it is also why a classical density embedded via $\Phi$ can formally saturate or violate covariance-based bounds without thereby being a quantum state, a point we return to for entanglement criteria below.

Having fixed the kinematics in both languages, as seen in Table \ref{tab:compare} we can now compare how this algebraic structure drives time evolution: Liouville's equation versus von Neumann evolution and their phase-space forms in terms of Poisson versus Moyal brackets.

\section{Time Evolution}
\label{sec:time_evolution}
Once the multiplication rule is fixed, i.e. pointwise for classical symbols, $\star$-product for quantum symbols, the corresponding brackets and hence the dynamical generators are fixed as well. In other words, the representational translation does not create a new dynamical principle, but it relocates the same structural difference into the bracket that generates time evolution. We have Poisson flow versus Moyal flow in phase space, and their Hilbert-space counterparts under the inverse map. This product structure allows us to immediately write down the Liouville--type equations for a Hamiltonian $H$ \cite{goldstein_klassische_2012}
\begin{equation}
\label{eq:Liouville}
    \partial_t f(q,p,t) = \{ H, f(q,p,t) \},
\end{equation}
where the quantum counterpart simply replaces the Poisson bracket with a Moyal bracket
\begin{equation}
\label{eq:quantum_Liouville}
    \partial_t W(q,p,t) = \{ H, W(q,p,t) \}_\text{MB}.
\end{equation}

For Hamiltonians at most quadratic in $(q,p)$, the Moyal bracket equals the Poisson bracket exactly, hence Liouville and Wigner evolution coincide. This quantum PS evolution is also central to a broad class of practical semiclassical and mixed quantum--classical schemes, because it provides a controlled route to approximations obtained by expanding the $\hbar$-dependent quantum corrections around the classical Poisson bracket and then keeping only the leading terms (e.g., linearized/classical--Wigner or truncated--Wigner--type dynamics). In particular, such truncations underpin widely used semiclassical PS methods and their systematic analyses~\cite{polkovnikov_phase_2010}, and they form the conceptual and computational basis of several approaches in chemical physics for time--correlation functions, spectroscopy, and molecular dynamics in high--dimensional systems~\cite{heller_wigner_1976, liu_linearized_2007, miller_perspective_2012}. This is a prime application of a controlled truncation of the $\star$--deformation.

Apart from various applications of these equations, it can be instructive to transform them to their HS form to find both similarities and differences in the language of quantum theory. Naturally, from Eq. \eqref{eq:comm_MB}, it is immediately clear that Eq. \eqref{eq:quantum_Liouville} will give the well--known von Neumann (vN) equation
\begin{equation}
\label{eq:vN}
    i \hbar \frac{\partial \hat{\rho}}{\partial t} = [H, \hat{\rho}].
\end{equation}
More interestingly, however, is the classical counterpart of this equation, which dates back to Bohm \& Hiley~\cite{bohm_quantum_1981} (BH in Table \ref{tab:compare}), where the Weyl transform of Eq. \eqref{eq:Liouville} gives
\begin{equation}
\begin{split}
    i\hbar \partial_t\rho_f(x,y,t) = - \frac{\hbar^2}{2m}\left( \frac{\partial^2}{\partial x^2} - \frac{\partial^2}{\partial y^2} \right) \rho_f \\ + (x-y) V' \left(\frac{x+y}{2}  \right) \rho_f,
\end{split}
\end{equation}
in matrix elements $\rho_f(x,y,t) = \bra{x}\hat{\rho}\ket{y}$. Note that if one were to also express the vN equation \eqref{eq:vN} in matrix elements, the potential term would be $V(x)-V(y)$. An important consequence of this is that \textit{up to quadratic order in the potential}, both classical and quantum time \textit{evolution in HS coincide exactly}. To see this more clearly, take, for instance, the example of a harmonic oscillator
\begin{equation}
H(q,p)=\frac{p^2}{2m}+\frac{1}{2}m\omega^2 q^2 .
\end{equation}
Because $H$ is at most quadratic in $(q,p)$, the Moyal bracket reduces \emph{exactly} to the Poisson bracket,
\begin{equation}
\{H,W\}_{MB}=\{H,W\}_{PB},
\end{equation}
so Eqs.~\eqref{eq:Liouville} and \eqref{eq:quantum_Liouville} generate the \emph{same} phase-space flow for both $f$ and $W$. Concretely, the solution is just rigid rotation in phase space
\begin{align}
W(q,p,t) &= W_0\!\left(q\cos\omega t-\frac{p}{m\omega}\sin\omega t,\; p\cos\omega t+m\omega q\sin\omega t\right), \\
f(q,p,t) &= f_0\!\left(q\cos\omega t-\frac{p}{m\omega}\sin\omega t,\; p\cos\omega t+m\omega q\sin\omega t\right).
\end{align}
In particular, a nonnegative Gaussian Wigner function remains nonnegative and Gaussian, simply rotated, illustrating in the most transparent setting that the WW translation does not introduce a new dynamical principle, as for quadratic generators, the $\star$--deformation does not contribute, and classical and quantum predictions coincide. This coincidence is not a representational artifact but a structural statement: up to quadratic order, the $\star$--deformation does not contribute, and classical and quantum flows agree. Beyond that order, the deformation becomes operationally relevant, and the two evolutions diverge. One should not overemphasize this coincidence however. Although the state and its evolution agree for quadratic $H$, the difference can resurface at the level of measurement, where the discrete spectrum and zero-point energy of the oscillator only appear upon measuring energy, an observable whose spectrum is discrete unlike its classical counterpart. As discussed in Sec.~\ref{sec:measurement}, even phase-space measurements distinguish the two theories through the post-measurement disturbance dictated by the uncertainty principle, whereby a sharp position readout leaves the momentum uniform [cf.\ Eq.~\eqref{eq:pos-eigen-wigner}].

To see where the deformation becomes operationally relevant, take a standard non-quadratic Hamiltonian
\begin{equation}
H(q,p)=\frac{p^2}{2m}+V(q),
\end{equation}
and expand the Moyal bracket in $\hbar$. Using the $\star$--product expansion in Eq. \eqref{eq:star_expansion}, the quantum evolution can be written as a classical Liouville term plus $\hbar$-dependent corrections~\cite{hillery_distribution_1984, polkovnikov_phase_2010, curtright_quantum_2012}
\begin{equation}
\begin{split}
\partial_t W \;=\; &-\frac{p}{m}\,\partial_q W \;+\; V'(q)\,\partial_p W \\&\;-\;\frac{\hbar^2}{24}\,V^{(3)}(q)\,\partial_p^{3}W \;+\; O(\hbar^4).
\end{split}
\label{eq:moyal_expansion_potential}
\end{equation}
The key point is that the first genuinely quantum term is proportional to $V^{(3)}(q)$, which vanishes for quadratic $V$, while for cubic and higher potentials it does not. This directly matches (in dynamical form) the earlier Groenewold--van Hove message that obstructions first arise beyond quadratic order.

As a concrete illustration, for the quartic oscillator $V(q)=\lambda q^4$ one has $V'(q)=4\lambda q^3$ and $V^{(3)}(q)=24\lambda q$, so Eq.~\eqref{eq:moyal_expansion_potential} becomes
\begin{equation}
\partial_t W \;=\; -\frac{p}{m}\,\partial_q W \;+\; 4\lambda q^3\,\partial_p W \;-\; \hbar^2 \lambda q \,\partial_p^{3}W \;+\; O(\hbar^4),
\end{equation}
making explicit how quantum evolution deviates from classical phase-space flow once the Hamiltonian probes beyond quadratic structure.

This could, for instance, be used to figure out how fast and how much classical and quantum states  diverge with respect to time for potentials of cubic order and beyond. A critical example of this would be current proposals for witnessing gravity--mediated entanglement \cite{bose_spin_2017, marletto_gravitationally_2017, krisnanda_observable_2020}, where a clear distinction between classical and quantum evolution is of utmost importance.

\section{Measurement}
\label{sec:measurement}
Additionally, the measurement to which we will now refer reveals a substantial disparity between classical and quantum mechanics. Operationally, a measurement produces a classical outcome $m$ (a record that can be copied, stored, and broadcast). In both classical and quantum theories, the minimal specification of this is an \emph{outcome probability rule}, implemented in classical mechanics via response functions $\xi(m|q,p)$ and in quantum mechanics via the Born rule for a POVM $\{E_m\}_m$. In many physical implementations, the same measurement interaction also induces a conditional transformation of the system state. This additional information is captured by a (classical or quantum) \emph{instrument}, denoting a family of outcome-labeled maps, which is not fixed by the probability rule alone. In what follows we therefore separate (i) outcome statistics from (ii) conditional state updates~\cite{davies_operational_1970, kraus_6_1983, luders_uber_1950}, and we use the WW correspondence to express both layers in a common language. Throughout, we treat outcome statistics as the defining content of a measurement and post-measurement states as part of the instrument. While this is definitional, keeping the two layers separate is what lets us localize the classical--quantum difference in the update rule rather than the probabilities. The key diagnostic is that outcome probabilities can be made to look overlap-like in both languages, whereas the conditioned update exposes the noncommutative structure. The WW correspondence allows us to express both classical and quantum measurement statistics in either PS or HS language. This reveals which differences are merely representational (e.g., negativity appearing under a map) and which reflect a genuine structural distinction, i.e., noncommutativity or the $\star$--product.

\subsection{Outcome statistics: response functions and POVMs}
In classical mechanics, a state is a probability density $f(q,p)\ge 0$ on phase space $\Gamma$, and a measurement is most naturally described by a \emph{response function} $\xi(m|q,p)$ giving the probability density of obtaining an outcome $m$ when the system is at $(q,p)$. Outcome probabilities are then obtained via
\begin{equation}
p(m)=\int_{\Gamma} dq\,dp\; f(q,p)\,\xi(m|q,p),
\label{eq:classical-outcome}
\end{equation}
Idealized sharp readouts correspond to distributions such as $\xi(q_0|q,p)\propto \delta(q-q_0)$ for a perfect position readout.

In quantum mechanics, the analogue of $\xi(m|q,p)$ is a POVM $\{E_m\}_m$, $E_m\ge 0$, $\sum_m E_m = \mathbb{I}$, yielding the Born rule \cite{holevo_probabilistic_2011}
\begin{equation}
p(m)=\text{Tr}(\hat\rho\,E_m).
\label{eq:born}
\end{equation}

The WW correspondence makes the probability rule look formally similar in phase space for both theories. Let $W_\rho(q,p)$ be the Wigner function of  $\hat\rho$, and let $E_m^W(q,p)$ denote the Weyl symbol of the POVM $E_m$. Using the trace relation, we obtain the quantum probability in phase space
\begin{equation}
p(m)=\mathrm{Tr}(\hat\rho\,E_m)
     = \int_{\Gamma} dq\,dp\; W_\rho(q,p)\,E_m^W(q,p).
\label{eq:wigner-born}
\end{equation}
Thus, quantum outcome probabilities appear as an \emph{overlap integral} between the state symbol and the effect symbol, similarly to Eq. \eqref{eq:classical-outcome}.

For classical states embedded into Hilbert space via Weyl quantization, $\hat\rho_f=\Phi[f]$, the same trace formula yields
\begin{equation}
\mathrm{Tr}(\hat\rho_f E_m)=\int dq\,dp\; f(q,p)\,E_m^W(q,p),
\label{eq:classical-hs-formal}
\end{equation}
which is \emph{not guaranteed to be nonnegative} for arbitrary $E_m$ because $\hat\rho_f$ need not be positive and, equivalently, $E_m^W$ need not be a genuine classical response function. This is the precise sense in which a naive “Born rule on $\hat\rho_f$'' is only a \emph{formal} measurement rule for embedded classical states, as it can output negative “probabilities” unless one restricts to an operationally admissible subset of effects. Operationally admissible effects are those that yield $\text{Tr}[\hat \rho_f E_m] \geq 0$ for all $f$. In phase space, this would correspond to effects whose Weyl symbols behave like genuine response functions, i.e., sufficient conditions would be that they are nonnegative and normalized.

\subsection{Conditional updates: instruments and $\star$-deformation}
In classical mechanics, the conditioned state is obtained by Bayes’ rule~\cite{cox1946probability}
\begin{equation}
f_m(q,p)=\frac{f(q,p)\,\xi(m|q,p)}{p(m)}.
\label{eq:bayes-update}
\end{equation}
Crucially, classical conditioning does not require any notion of projection in Hilbert space, because it is a purely probabilistic update. Any physical disturbance of the system can be included by upgrading the description to an \emph{instrument}, i.e.\ a family of transition kernels $T_m(q',p'|q,p)$ (one for each outcome) such that the unnormalized posterior becomes \cite{davies_operational_1970}
\begin{equation}
f_m(q',p')=\int dq\,dp\; T_m(q',p'|q,p)\,f(q,p),
\end{equation}
with $p(m)=\int dq'\,dp'\, f_m(q',p')$.

In quantum mechanics, the post-measurement state is not determined by the POVM alone but by a measurement \emph{instrument}. In Kraus form \cite{kraus_6_1983},
\begin{equation}
\rho_m = \frac{\sum_\alpha K_{m,\alpha}\,\hat\rho\,K_{m,\alpha}^\dagger}{p(m)},
\qquad
E_m=\sum_\alpha K_{m,\alpha}^\dagger K_{m,\alpha}.
\label{eq:kraus-update}
\end{equation}
Projective measurements are recovered as the special case $K_m = \Pi_m$ (orthogonal projectors), for which $\hat\rho_m = \Pi_m\hat\rho\Pi_m/p(m)$ \cite{luders_uber_1950}.

The deeper distinction is that quantum kinematics ties measurement updates to the noncommutative product structure. In the WW language, operator multiplication is represented by the Moyal $\star$--product. Consequently, while the \emph{outcome probabilities} take the simple overlap form in Eq.~\eqref{eq:wigner-born}, the \emph{conditioned quantum state} generally involves $\star$--products when expressed in phase space. For an instrument with Kraus operators $K_{m,\alpha}$ and symbols $K_{m,\alpha}^W$, the unnormalized post-measurement Wigner function is schematically
\begin{equation}
\begin{split}
&\widetilde{W}_m \;\propto\; \sum_\alpha
\left(K_{m,\alpha}^W \star W_\rho \star (K_{m,\alpha}^W)^\ast\right),
\\
&W_m=\widetilde{W}_m/p(m),
\end{split}
\label{eq:wigner-update-star}
\end{equation}
highlighting that quantum conditioning is not a purely pointwise reweighting (Bayes rule), but a deformation driven by the $\star$--product and hence by noncommutativity. Only in special cases (e.g., coarse-grained measurements \cite{bibak2025classical}) does the $\star$--update reduce to a Bayes-like reweighting. An example for this can be seen in Appendix \ref{A:Bayes}.

Another example that makes the two update rules explicit is the simple example of a sharp readout, i.e. a measurement of position with the outcome $q=x_0$. We compare the classical and quantum instruments and find that although the outcome statistics are identical in form, the postmeasurement states differ. Classically the sharp position readout is the response function $\xi(x_0|q,p)$, and Bayes' rule \eqref{eq:bayes-update} gives 
\begin{equation}
    f_{x_0}(q,p)=\delta(q-x_0)\,\frac{f(x_0,p)}{f_Q(x_0)},
\qquad f_Q(x_0)=\int dp\, f(x_0,p).
\end{equation}
This pointwise reweighting sharpens $q$ to $x_0$ but acts as the identity on the momentum sector. The conditioned momentum distribution is the slice $f(x_0, p)/f_Q(x_0)$ and for an uncorrelated state $f=f_Q(q)f_P(p)$ it is simply $f_P(p)$, i.e. the momentum is left untouched. Quantum mechanically on the other hand, the same outcome is the projector $\hat\Pi_{x_0}=\ket{x_0}\bra{x_0}$. Its Weyl symbol is $\Pi_{x_0}^W(q,p)=\delta(q-x_0)$, identical to the classical response function, so by Eq.~\eqref{eq:wigner-born} the outcome probability
\begin{equation}
p(x_0)=\int dp\, W_\rho(x_0,p)
\end{equation}
is the position marginal of the Wigner function, mirroring the classical $f_Q(x_0)$ exactly. At the level of outcome statistics the two theories are indistinguishable in this representation. However, the postmeasurement state is where they part, as the projected state $\hat\rho_{x_0}=\ket{x_0}\bra{x_0}$ has Wigner function
\begin{equation}
\label{eq:pos-eigen-wigner}
W_{x_0}(q,p)=\frac{1}{2\pi\hbar}\,\delta(q-x_0),
\end{equation}
which is sharp in $q$ but uniform in $p$, delocalizing the momentum maximally. In PS this behaviour is produced by the $\star$-products of Eq.~\eqref{eq:wigner-update-star}, whose $p$ derivatives act on $W_\rho$ and inject momentum spread. The mechanism is quantified by regularizing the projector as the Gaussian effect $E_\sigma\propto\exp[-(\hat q-x_0)^2/2\sigma^2]$ of Appendix~\ref{A:Bayes}. Here we can see that a position measurement of resolution $\sigma$ imprints a momentum disturbance of order $\hbar/2\sigma$. For a sharp readout $\sigma \to 0$ it diverges, recovering the uniform in $p$ Wigner function \eqref{eq:pos-eigen-wigner}. Classically the same Gaussian response $\xi\propto\exp[-(q-x_0)^2/2\sigma^2]$ multiplies only the $q$ dependence of $f$ for every $\sigma$ and no momentum disturbance is ever generated.

The WW correspondence therefore separates two conceptually distinct issues.
First, \emph{negativity} can arise on either side of the translation (Wigner negativity for quantum states \cite{ferrie_quasi-probability_2011} or non-positivity of $\rho_f$ for embedded classical states), and this feature by itself is representational, as it depends on how states are embedded and which effects are deemed admissible.
Second, the \emph{structural} difference appears at the level of conditioning: classical conditioning is pointwise reweighting (Bayes), whereas quantum conditioning is implemented by an instrument and is constrained by noncommutative operator multiplication.
In the WW language, this constraint becomes explicit because the conditional update generically involves $\star$-products of\ Eq.~\eqref{eq:wigner-update-star}.
Thus, rather than viewing ``collapse'' as an additional postulate beyond the probability rule, one may view it more precisely as the theory's rule for \emph{state update conditioned on a recorded outcome}; the WW formalism highlights that its non-Bayesian character is directly tied to the underlying noncommutative algebra. This is not to deny disturbance, but rather, it isolates the structural reason the update cannot be purely Bayesian in general.

\section{Entanglement}
\label{sec:entanglement}
Since this article focuses on the Wigner--Weyl (WW) framework as a common language for classical and quantum physics, it is useful to briefly summarize how entanglement fits into this representational perspective.
In Ref.~\cite{schlegel_entanglement_2025} we have studied bipartite correlations when classical PS distributions are mapped into HS via the Weyl transform (and, conversely, quantum states are represented in PS via the Wigner transform). This comparison separates features that are merely representational from those that mark genuine physical differences. See also Ref.~\cite{khalid2025classical} for a related classical analogue of entanglement, defined through the phase-space mutual information of subsystems in the kicked top.

Recall from a previous section (in particular, Eqs. \eqref{eq:def_Q} and \eqref{eq:def_C}) that we can define two sets of states: (i) classical operators $\mathcal{C}$, i.e.\ trace--one operators whose Wigner representation is everywhere nonnegative, and (ii) quantum states $\mathcal{Q}$, i.e.\ positive semidefinite density operators, as seen in Fig. \ref{fig:WW}.
Neither set contains the other; instead, they overlap in $\mathcal{C}\cap \mathcal{Q}$, which includes all Wigner-positive quantum states (Gaussian states and, more generally, Wigner-positive mixed states \cite{brocker_mixed_1995}). Within this overlap, quadrature statistics alone do not determine whether the underlying preparation is classical or quantum, because the same nonnegative PS function can be interpreted as a classical probability density or as a quantum Wigner function.

\begin{figure}[h]
    \centering
    \includegraphics[width=1.0\linewidth]{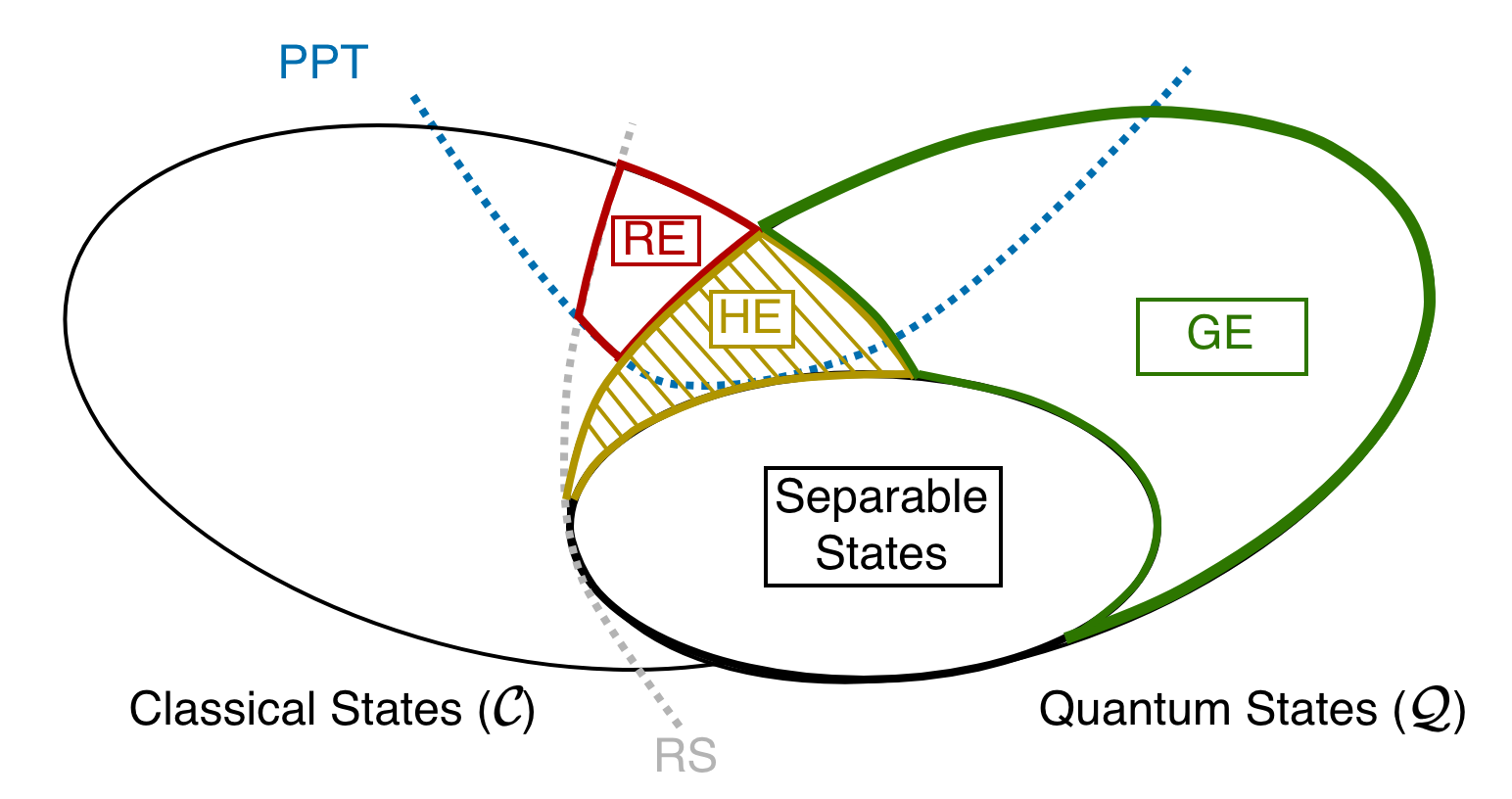}
    \caption{The entanglement hierarchy of Ref.~\cite{schlegel_entanglement_2025}
    located on the classical/quantum state spaces of Fig.~\ref{fig:WW}.
    \emph{Representational entanglement} (RE): a classical density $f\ge0$ whose
    Hilbert-space image $\hat\rho_f$ is not positive, so covariance witnesses are
    applied outside their domain. \emph{Hybrid entanglement} (HE): states that are
    positive and entangled yet Wigner-positive, reproducible by a classical
    phase-space model at the quadrature level ($\mathcal{C}\cap\mathcal{Q}$).
    \emph{Genuine entanglement} (GE): positive, entangled, and Wigner-negative,
    admitting no classical phase-space description.}
    \label{fig:entanglement}
\end{figure}

One can then introduce an operational hierarchy of entanglement--like behavior that arises when the observer has restricted access (Fig.~\ref{fig:entanglement}):
\begin{enumerate}
\item \emph{Representational entanglement (RE):} under limited data (e.g.,\ only second moments), classical mixtures can violate standard covariance-based entanglement criteria, such as the positive-partial-transpose (PPT) and Duan--Simon criteria \cite{adesso_entanglement_2007, duan_inseparability_2000, serafini_quantum_2023, simon_peres-horodecki_2000}, even though the associated Hilbert-space operator is not positive.
The apparent entanglement is then a representational artifact because it arises from applying quantum separability diagnostics outside their domain of validity when positivity is not guaranteed. Note that the correlations themselves are physically real, being ordinary statistical correlations of a classical joint distribution. What is representational is its diagnosis as entanglement, since criteria of this type are derived under the assumption $\hat\rho\ge0$ and are applied here to states in $\mathcal{C}\setminus\mathcal{Q}$, where that assumption does not hold.
\item \emph{Hybrid entanglement (HE):} if the restriction is lifted enough to test positivity (e.g.,\ via local homodyne tomography), one can identify states that are both positive and entangled while still admitting an everywhere nonnegative Wigner representation.
These states inhabit the overlap $\mathcal{C}\cap \mathcal{Q}$: they are entangled in Hilbert space, yet remain reproducible by a classical phase-space model under the same quadrature-level operational access.
\item \emph{Genuine entanglement (GE):} finally, states that are positive and entangled but exhibit Wigner-function negativity (or, equivalently, permit operational tests that go beyond jointly measurable phase-space observables) are not reproducible by any classical phase-space description.
This regime corresponds to entanglement accompanied by a nonclassicality signature.
\end{enumerate}

This hierarchy was made explicit by considering examples within these different regimes. A tunable non--Gaussian classical mixture (negative Hilbert space operator) of displaced Gaussians was used to violate the PPT criterion while satisfying covariance--based uncertainty relations, giving an example of RE. To visualize the HE regime and the crossing over into GE, we mixed $\ket{0}\bra{0}$ and $\ket{1}\bra{1}$, while inserting vacuum in the other mode. After applying a balanced beamsplitter, one obtains an entangled state that remains Wigner--positive for a finite parameter range while being non--Gaussian and only crossing over into the GE regime once the mixing weight for $\ket{1}\bra{1}$ was sufficiently large. We refer to Ref. \cite{schlegel_entanglement_2025} for the full analysis. Note that Bell inequalities cannot be violated by these classical systems. Every classical measurement is a response function $\xi(m|q,p)$ on phase space, so any collection of them is jointly measurable by a single readout of $(q,p)$, while the density $f(q,p)$ gives a global joint probability distribution for the outcomes of all of them at the same time. By Fine's theorem~\cite{fine_hidden_1982}, the existence of such a joint distribution is equivalent to the satisfaction of all Bell inequalities of the corresponding scenario. Thus a violation would require observables that are not jointly measurable~\cite{wolf_measurements_2009}, which a classical phase-space model does not possess. This should not be read as saying that Wigner-positive entangled states are Bell-local. The two-mode squeezed vacuum is Gaussian, hence lies in $\mathcal{C}\cap \mathcal{Q}$, yet violates a Bell inequality under displaced parity measurements \cite{PhysRevA.58.4345}. There is no tension, because the Weyl symbols of the parity effects are not nonnegative response functions, so such measurements fall outside the jointly measurable phase-space observables to which the classical model is restricted. 

From the WW standpoint, ``entanglement'' is not a single-layer diagnostic, as under restricted operational access, classical correlations may appear entangled (RE), while even genuine entangled quantum states may remain classically reproducible at the level of phase-space statistics (HE).
Only when entanglement is accompanied by a nonclassicality witness such as Wigner negativity (or equivalent operational tests beyond compatible PS observables) does it enter the genuine quantum regime (GE).

\section{Conclusion}
The central message of this work is that the physical content of a theory, meaning its outcome statistics, state updates and dynamics, is not tied to a preferred mathematical formalism. The Wigner–Weyl correspondence makes this concrete by providing an explicit and invertible two-way translation between Hilbert-space and phase-space descriptions for both classical and quantum mechanics, under which every prediction is preserved. This is not to say that all representations are equally convenient---different representations can make different structural features of a theory more or less transparent. In this pragmatic sense, a representation may appear natural when its defining constraints take a particularly simple form, like pointwise positivity of $f$ in phase space for classical mechanics, and operator positivity of $\hat \rho$ in Hilbert space for quantum mechanics. Yet this simplicity is a property of representation and not one of the physical content itself. A constraint that is transparent in one language, such as $\hat \rho \geq 0$, may become an unwieldy family of conditions in another, such as the KLM conditions on its Wigner function $W$. Conversely, moving to a less familiar representation can expose structural features that are obscured in the conventional formalism. The resulting mismatch under translation is the origin of the representational artifacts discussed above. While the mathematical language used to formulate a theory can be relevant, the representation dependent simplicity or complexity should not be mistaken for the physical boundary between the theories.

Viewed through this lens, several familiar ``quantum signatures'' are best understood as representational artifacts tied to positivity constraints rather than as intrinsic diagnostics. Because the WW maps are invertible on symbols/operators but do not preserve positivity, negativity can appear either as Wigner-function negativity for quantum states or as non-positivity of Hilbert-space images of classical distributions. Read operationally, such features signal incompatibility with a chosen admissible set of states/effects under a chosen representation, not a representation-independent boundary.

What \emph{does} survive a change of language is the structural distinction: noncommutativity. In phase space this appears as the $\hbar$-dependent $\star$--product and the associated Moyal bracket, and it is this deformation that underwrites incompatibility and constrains both dynamics and measurement beyond what classical Poisson/Bayes structure can reproduce. The same perspective clarifies when classical and quantum predictions coincide (e.g., quadratic Hamiltonians, where the deformation does not contribute) and where they must diverge (beyond quadratic order, where $\star$-corrections become dynamically relevant).

Extending the comparison to measurement sharpens the operational payoff of the representational viewpoint. While outcome probabilities can be written as overlap-type integrals in either language, the conditioned update rule exposes the genuine difference: classical conditioning is pointwise Bayesian reweighting, whereas quantum conditioning is generically $\star$--deformed due to noncommutative multiplication. In this sense, what appears as ``collapse'' in Hilbert space is, in the WW language, the manifestation of the same algebraic structure that distinguishes quantum from classical mechanics across representations.

Finally, when considering entanglement within this formalism, a rich hierarchy of nonseparability is revealed, separated by positivity constraints. Purely classical distributions (negative HS operators) can violate certain entanglement criteria for a restricted observer, and even when explicit state tomography is allowed, a hybrid regime remains. In this hybrid regime, it is unclear whether correlations are sourced by a classical or a quantum system, unless explicit checks of incompatibility or Wigner--negativity are supplemented, which would certify genuine entanglement. 

We can conclude that many features, like the states themselves, evolution, and entanglement, have closely related classical counterparts once representational biases are removed and the theories are compared on equal footing. Finding the representation--independent differences between the two theories, either for the state--update or the kinematics, requires the non--commutative structure of quantum mechanics, exemplified in phase space by the Moyal $\star$--product. It is important to note here that differences in the kinematic relations, as well as in the time evolution, only emerge when quadrature variables of cubic or higher order are considered. This common theme is emerging throughout this comparison and has important implications for any claim of nonclassicality. Both in theory and in experiment, one needs to make sure that the system of interest exhibits either some form of incompatibility or a negative Wigner function. More generally, we can see that just the use of a Hilbert--space language to describe our system must not be confused with a genuine signature of ``quantumness.''

\emph{Acknowledgments.--}We warmly thank Nicolas Gisin, Tomasz Paterek, and Ankit Kumar for their insightful comments and helpful discussions on this work. This research was funded in whole, or in part, by the Austrian Science Fund (FWF) [10.55776/F71] and
[10.55776/P36994]. For open access purposes, the author(s) has applied a CC BY public copyright license to any author accepted manuscript version arising from this submission.


\appendix
\setcounter{secnumdepth}{3}
\section{Recovering Bayes reweighting}
\label{A:Bayes}
To see how Lüders rule in phase space
\begin{equation}
\label{eq:A_lüder}
    {W}_m \;= \; \frac{1}{p(m)} \sum_\alpha
\left(K_{m,\alpha}^W \star W_\rho \star (K_{m,\alpha}^W)^\ast\right),
\end{equation}
recovers the Bayes reweighting of Eq. \eqref{eq:bayes-update}, consider, for instance, a Gaussian POVM element of the form 
\begin{equation}
\label{eq:A_POVM}
    E_m \propto \exp \left[ -\frac{(\hat{q}-q_m)^2}{2\sigma^2} \right],
\end{equation}
whose corresponding PS symbol $E_m^W$ is also a positive Gaussian in $q$ and its Kraus operator is given by $K^W = (E_m^W)^{1/2}$. Recall that more generally, the Moyal product of a Weyl symbol $A$ that only carries $q$--dependence and a symbol $B$ that carries both $q$ and $p$ dependence is given by
\begin{equation}
    A \star B = \sum_{n=0}^\infty \frac{1}{n!} \left( \frac{i\hbar}{2} \right)^n A^{(n)}(q) \partial_p^n B(q,p),
\end{equation}
which will in general not reduce to a pointwise product, because higher $p$--derivatives appear on B, which are weighted by $A^{(n)}(q)$. However, if we consider now, for instance, the Gaussian POVM of Eq. \eqref{eq:A_POVM}, we find that $A$ varies on the length--scale $\sigma$ and we know that $A^{(n)}/A \sim \sigma^{-n}$. In a similar vein, our Wigner function varies on the scale $\Delta p$, giving $(\partial_p^nB)/B \sim (\Delta p)^{-n}$. This allows us to make the scaling identification
\begin{equation}
    \left| \frac{ A^{(n)}\partial_p^n B}{AB} \right| \sim \left( \frac{\hbar}{2\sigma \Delta p} \right)^n,
\end{equation}
yielding a handy condition: if $\sigma \Delta p \gg \hbar$, then $A \star B \approx AB$. This same suppression factor can also be found in the $\star$-commutator
\begin{equation}
    A \star B - B \star A = i \hbar \{A, B\} + \mathcal{O}(\hbar^3) = i \hbar A'(q) \partial_p B(q,p) + \mathcal{O}(\hbar^3),
\end{equation}
giving the same coarse--graining condition $\sigma \Delta p \gg \hbar$. Now assume that this condition holds true. Plugging into Lüders rule \eqref{eq:A_lüder}, gives 
\begin{equation}
    W_m \propto A_m \star W \star A_m = A^2W + \frac{\hbar^2}{4}\left( (A')^2-AA'' \right) \partial_p^2W + \mathcal{O}(\hbar^4),
\end{equation}
where we see that due to this coarse--graining, the leading term dominates, and plugging in $A=K^W$ yields
\begin{equation}
    A \star W \star A \approx A^2 W = (K^W)^2 W = E^W W. 
\end{equation}
After renormalization we can therefore recover Bayes reweighting 
\begin{equation}
    W_m \approx \frac{E_m^W W }{\int dq \, dp \, W(q,p) \, E^W(q,p)},
\end{equation}
akin to Eq. \eqref{eq:bayes-update}.



\begin{thebibliography}{10}

\bibitem{moyal_quantum_1949}
J.~E. Moyal.
\newblock Quantum mechanics as a statistical theory.
\newblock {\em Mathematical Proceedings of the Cambridge Philosophical
  Society}, 45(1):99--124, January 1949.

\bibitem{wigner_quantum_1932}
E.~Wigner.
\newblock On the {Quantum} {Correction} {For} {Thermodynamic} {Equilibrium}.
\newblock {\em Physical Review}, 40(5):749--759, June 1932.

\bibitem{weyl_quantenmechanik_1927}
H.~Weyl.
\newblock Quantenmechanik und {Gruppentheorie}.
\newblock {\em Zeitschrift für Physik}, 46(1):1--46, November 1927.

\bibitem{groenewold_principles_1946}
Hilbrand~Johannes Groenewold.
\newblock On the principles of elementary quantum mechanics.
\newblock {\em Physica}, 12(7):405--460, 1946.

\bibitem{van_hove_sur_1951}
Léon Van~Hove.
\newblock Sur le problème des relations entre les transformations unitaires de
  la {Mécanique} quantique et les transformations canoniques de la
  {Mécanique} classique.
\newblock {\em Bulletins de l'Académie Royale de Belgique}, 37(1):610--620,
  1951.

\bibitem{baker_formulation_1958}
George~A. Baker.
\newblock Formulation of quantum mechanics based on the quasi-probability
  distribution induced on phase space.
\newblock {\em Physical Review}, 109(6):2198--2206, 1958.

\bibitem{bohm_quantum_1981}
D.~Bohm and B.~J. Hiley.
\newblock On a quantum algebraic approach to a generalized phase space.
\newblock {\em Foundations of Physics}, 11(3):179--203, April 1981.

\bibitem{curtright_quantum_2012}
Thomas~L. Curtright and Cosmas~K. Zachos.
\newblock Quantum {Mechanics} in {Phase} {Space}.
\newblock {\em Asia Pacific Physics Newsletter}, 01(01):37--46, May 2012.

\bibitem{bartlett_reconstruction_2012}
Stephen~D. Bartlett, Terry Rudolph, and Robert~W. Spekkens.
\newblock Reconstruction of {Gaussian} quantum mechanics from {Liouville}
  mechanics with an epistemic restriction.
\newblock {\em Physical Review A}, 86(1):012103, July 2012.
\newblock arXiv:1111.5057 [quant-ph].

\bibitem{spekkens_defense_2007}
Robert~W. Spekkens.
\newblock Evidence for the epistemic view of quantum states: {A} toy theory.
\newblock {\em Physical Review A}, 75(3):032110, March 2007.
\newblock arXiv:quant-ph/0401052.

\bibitem{santo_physics_2019}
Flavio~Del Santo and Nicolas Gisin.
\newblock Physics without {Determinism}: {Alternative} {Interpretations} of
  {Classical} {Physics}.
\newblock {\em Physical Review A}, 100(6):062107, December 2019.
\newblock arXiv:1909.03697 [quant-ph].

\bibitem{santo_which_2025}
Flavio~Del Santo and Nicolas Gisin.
\newblock Which features of quantum physics are not fundamentally quantum but
  are due to indeterminism?
\newblock {\em Quantum}, 9:1686, April 2025.
\newblock arXiv:2409.10601 [quant-ph].

\bibitem{schlegel_entanglement_2025}
Samuel Schlegel, Borivoje Dakić, and Flavio~Del Santo.
\newblock Entanglement without {Quantum} {Mechanics}: {Operational}
  {Constraints} on the {Quantum} {Signature}.
\newblock {\em Physical Review Research}, 8:033313, 2026.
\newblock arXiv:2512.14834 [quant-ph].

\bibitem{gerry_introductory_2023}
Christopher~C. Gerry and Peter~L. Knight.
\newblock {\em Introductory {Quantum} {Optics}}.
\newblock Cambridge University Press, November 2023.

\bibitem{case_wigner_2008}
William~B. Case.
\newblock Wigner functions and {Weyl} transforms for pedestrians.
\newblock {\em American Journal of Physics}, 76(10):937--946, October 2008.

\bibitem{hillery_distribution_1984}
M.~Hillery, R.~F. O'Connell, M.~O. Scully, and E.~P. Wigner.
\newblock Distribution functions in physics: {Fundamentals}.
\newblock {\em Physics Reports}, 106(3):121--167, April 1984.

\bibitem{mauro_topics_2003}
D.~Mauro.
\newblock {\em Topics in {Koopman}--von {Neumann} {Theory}}.
\newblock PhD thesis, Universit\`a degli Studi di Trieste, 2002.
\newblock arXiv:quant-ph/0301172.

\bibitem{morgan_algebraic_2020}
Peter Morgan.
\newblock An algebraic approach to {Koopman} classical mechanics.
\newblock {\em Annals of Physics}, 414:168090, 2020.

\bibitem{galazo_classical_2020}
Raquel Galazo, Irene Bartolom\'e, Laura Ares, and Alfredo Luis.
\newblock Classical and quantum complementarity.
\newblock {\em Physics Letters A}, 384:126849, 2020.

\bibitem{Narcowich1989}
Francis~J. Narcowich.
\newblock Distributions of {$\hbar$}-positive type and applications.
\newblock {\em Journal of Mathematical Physics}, 30(11):2565--2573, November
  1989.

\bibitem{kastler1965c}
Daniel Kastler.
\newblock The c*-algebras of a free boson field: I. discussion of the basic
  facts.
\newblock {\em Communications in Mathematical Physics}, 1(1):14--48, 1965.

\bibitem{Loupias_Miracle_1966}
G.~Loupias and S.~Miracle-Sole.
\newblock C*-algèbres des systèmes canoniques. {I}.
\newblock {\em Communications in Mathematical Physics}, 2(1):31--48, 1966.

\bibitem{Bochner1933}
S.~Bochner.
\newblock Monotone {Funktionen}, {Stieltjessche} {Integrale} und harmonische
  {Analyse}.
\newblock {\em Mathematische Annalen}, 108(1):378--410, 1933.

\bibitem{narcowich_necessary_1986}
Francis~J. Narcowich and R.~F. O’Connell.
\newblock Necessary and sufficient conditions for a phase-space function to be
  a {Wigner} distribution.
\newblock {\em Physical Review A}, 34(1):1--6, July 1986.

\bibitem{stratonovich1957distributions}
Ruslan~Leont’evich Stratonovich.
\newblock On distributions in representation space.
\newblock {\em Soviet Physics JETP}, 4(6):891--898, 1957.

\bibitem{kenfack_negativity_2004}
Anatole Kenfack and Karol \.Zyczkowski.
\newblock Negativity of the {Wigner} function as an indicator of
  nonclassicality.
\newblock {\em Journal of Optics B: Quantum and Semiclassical Optics},
  6(10):396--404, October 2004.
\newblock arXiv:quant-ph/0406015.

\bibitem{mandilara_extending_2009}
A.~Mandilara, E.~Karpov, and N.~J. Cerf.
\newblock Extending {Hudson}'s theorem to mixed quantum states.
\newblock {\em Physical Review A}, 79(6):062302, June 2009.

\bibitem{hudson_when_1974}
R.~L. Hudson.
\newblock When is the {Wigner} quasi-probability density non-negative?
\newblock {\em Reports on Mathematical Physics}, 6(2):249--252, October 1974.

\bibitem{CorderoDeGossonNicola2019}
Elena Cordero, Maurice de~Gosson, and Fabio Nicola.
\newblock On the positivity of trace class operators.
\newblock {\em Advances in Theoretical and Mathematical Physics},
  23(8):2061--2091, 2019.

\bibitem{DiasPrata2005}
Nelson~C. Dias and Jos{\'e}~N. Prata.
\newblock Deformation quantization and {Wigner} functions.
\newblock {\em Modern Physics Letters A}, 20:1371, 2005.
\newblock arXiv:hep-th/0504166.

\bibitem{serafini_quantum_2023}
Alessio Serafini.
\newblock {\em Quantum {Continuous} {Variables}: {A} {Primer} of {Theoretical}
  {Methods}}.
\newblock CRC Press, Boca Raton, 2 edition, August 2023.

\bibitem{weedbrook_gaussian_2012}
Christian Weedbrook, Stefano Pirandola, Raul Garcia-Patron, Nicolas~J. Cerf,
  Timothy~C. Ralph, Jeffrey~H. Shapiro, and Seth Lloyd.
\newblock Gaussian {Quantum} {Information}.
\newblock {\em Reviews of Modern Physics}, 84(2):621--669, May 2012.
\newblock arXiv:1110.3234 [quant-ph].

\bibitem{fujii_new_2004}
Kazuyuki Fujii and Tatsuo Suzuki.
\newblock A {New} {Symmetric} {Expression} of {Weyl} {Ordering}.
\newblock {\em Modern Physics Letters A}, 19(11):827--840, April 2004.
\newblock arXiv:quant-ph/0304094.

\bibitem{cahill_ordered_1969}
K.~E. Cahill and R.~J. Glauber.
\newblock Ordered {Expansions} in {Boson} {Amplitude} {Operators}.
\newblock {\em Physical Review}, 177(5):1857--1881, January 1969.

\bibitem{carosso_quantization_2022}
Andrea Carosso.
\newblock Quantization: {History} and {Problems}.
\newblock {\em Studies in History and Philosophy of Science}, 96:35, 2022.
\newblock arXiv:2202.07838 [physics].

\bibitem{bayen_deformation_1978_I}
F.~Bayen, M.~Flato, C.~Fronsdal, A.~Lichnerowicz, and D.~Sternheimer.
\newblock Deformation theory and quantization. {I}. deformations of symplectic
  structures.
\newblock {\em Annals of Physics}, 111(1):61--110, 1978.

\bibitem{bayen_deformation_1978_II}
F.~Bayen, M.~Flato, C.~Fronsdal, A.~Lichnerowicz, and D.~Sternheimer.
\newblock Deformation theory and quantization. {II}. physical applications.
\newblock {\em Annals of Physics}, 111(1):111--151, 1978.

\bibitem{goldstein_klassische_2012}
Herbert Goldstein, Charles P.~Poole~Jr., and John L.~Safko~Sr.
\newblock {\em Klassische {Mechanik}}.
\newblock Wiley-VCH, Weinheim, 3 edition, 2006.

\bibitem{polkovnikov_phase_2010}
Anatoli Polkovnikov.
\newblock Phase space representation of quantum dynamics.
\newblock {\em Annals of Physics}, 325(8):1790--1852, August 2010.
\newblock arXiv:0905.3384 [cond-mat].

\bibitem{heller_wigner_1976}
Eric~J. Heller.
\newblock Wigner phase space method: {Analysis} for semiclassical applications.
\newblock {\em The Journal of Chemical Physics}, 65(4):1289--1298, August 1976.

\bibitem{liu_linearized_2007}
Jian Liu and William~H. Miller.
\newblock Linearized semiclassical initial value time correlation functions
  using the thermal {Gaussian} approximation: {Applications} to condensed phase
  systems.
\newblock {\em The Journal of Chemical Physics}, 127(11):114506, September
  2007.

\bibitem{miller_perspective_2012}
William~H. Miller.
\newblock Perspective: {Quantum} or classical coherence?
\newblock {\em The Journal of Chemical Physics}, 136(21):210901, June 2012.

\bibitem{bose_spin_2017}
Sougato Bose, Anupam Mazumdar, Gavin~W. Morley, Hendrik Ulbricht, Marko Toroš,
  Mauro Paternostro, Andrew Geraci, Peter Barker, M.~S. Kim, and Gerard
  Milburn.
\newblock Spin {Entanglement} {Witness} for {Quantum} {Gravity}.
\newblock {\em Physical Review Letters}, 119(24):240401, December 2017.
\newblock arXiv:1707.06050.

\bibitem{marletto_gravitationally_2017}
C.~Marletto and V.~Vedral.
\newblock Gravitationally {Induced} {Entanglement} between {Two} {Massive}
  {Particles} is {Sufficient} {Evidence} of {Quantum} {Effects} in {Gravity}.
\newblock {\em Physical Review Letters}, 119(24):240402, December 2017.

\bibitem{krisnanda_observable_2020}
Tanjung Krisnanda, Guo~Yao Tham, Mauro Paternostro, and Tomasz Paterek.
\newblock Observable quantum entanglement due to gravity.
\newblock {\em npj Quantum Information}, 6:12, 2020.
\newblock arXiv:1906.08808.

\bibitem{davies_operational_1970}
E.~B. Davies and J.~T. Lewis.
\newblock An operational approach to quantum probability.
\newblock {\em Communications in Mathematical Physics}, 17(3):239--260,
  September 1970.

\bibitem{kraus_6_1983}
Karl Kraus.
\newblock Coexistent effects and observables.
\newblock In A.~Böhm, J.~D. Dollard, and W.~H. Wootters, editors, {\em States,
  {Effects}, and {Operations}: {Fundamental} {Notions} of {Quantum} {Theory}:
  {Lectures} in {Mathematical} {Physics} at the {University} of {Texas} at
  {Austin}}, volume 190 of {\em Lecture Notes in Physics}, pages 103--149.
  Springer, Berlin, Heidelberg, 1983.

\bibitem{luders_uber_1950}
Gerhart Lüders.
\newblock Über die {Zustandsänderung} durch den {Meßprozeß}.
\newblock {\em Annalen der Physik}, 443(5-8):322--328, 1950.

\bibitem{holevo_probabilistic_2011}
Alexander~S. Holevo.
\newblock {\em Probabilistic and {Statistical} {Aspects} of {Quantum}
  {Theory}}.
\newblock Edizioni della Normale, Pisa, 2 edition, 2011.

\bibitem{cox1946probability}
Richard~T Cox.
\newblock Probability, frequency and reasonable expectation.
\newblock {\em American journal of physics}, 14(1):1--13, 1946.

\bibitem{bibak2025classical}
Fatemeh Bibak, Carlo Cepollaro, Nicol{\'a}s~Medina S{\'a}nchez, Borivoje
  Daki{\'c}, and {\v{C}}aslav Brukner.
\newblock The classical limit of quantum mechanics through coarse-grained
  measurements.
\newblock {\em arXiv preprint arXiv:2503.15642}, 2025.

\bibitem{ferrie_quasi-probability_2011}
Christopher Ferrie.
\newblock Quasi-probability representations of quantum theory with applications
  to quantum information science.
\newblock {\em Reports on Progress in Physics}, 74(11):116001, October 2011.

\bibitem{khalid2025classical}
Bilal Khalid and Sabre Kais.
\newblock Classical analog of entanglement for a kicked top.
\newblock {\em Physical Review A}, 112(1):012201, 2025.

\bibitem{brocker_mixed_1995}
T.~Br{\"o}cker and R.~F. Werner.
\newblock Mixed states with positive {W}igner functions.
\newblock {\em Journal of Mathematical Physics}, 36(1):62--75, 1995.

\bibitem{adesso_entanglement_2007}
Gerardo Adesso and Fabrizio Illuminati.
\newblock Entanglement in continuous variable systems: {Recent} advances and
  current perspectives.
\newblock {\em Journal of Physics A: Mathematical and Theoretical},
  40(28):7821--7880, July 2007.
\newblock arXiv:quant-ph/0701221.

\bibitem{duan_inseparability_2000}
Lu-Ming Duan, G.~Giedke, J.~I. Cirac, and P.~Zoller.
\newblock Inseparability {Criterion} for {Continuous} {Variable} {Systems}.
\newblock {\em Physical Review Letters}, 84(12):2722--2725, March 2000.

\bibitem{simon_peres-horodecki_2000}
R.~Simon.
\newblock Peres-{Horodecki} separability criterion for continuous variable
  systems.
\newblock {\em Physical Review Letters}, 84(12):2726--2729, March 2000.
\newblock arXiv:quant-ph/9909044.

\bibitem{fine_hidden_1982}
Arthur Fine.
\newblock Hidden variables, joint probability, and the {Bell} inequalities.
\newblock {\em Physical Review Letters}, 48(5):291--295, 1982.

\bibitem{wolf_measurements_2009}
Michael~M. Wolf, David Perez-Garcia, and Carlos Fernandez.
\newblock Measurements {Incompatible} in {Quantum} {Theory} {Cannot} {Be}
  {Measured} {Jointly} in {Any} {Other} {No}-{Signaling} {Theory}.
\newblock {\em Physical Review Letters}, 103(23):230402, December 2009.

\bibitem{PhysRevA.58.4345}
Konrad Banaszek and Krzysztof W\'odkiewicz.
\newblock Nonlocality of the {Einstein-Podolsky-Rosen} state in the {Wigner}
  representation.
\newblock {\em Phys. Rev. A}, 58:4345--4347, Dec 1998.

\end{thebibliography}
\end{document}